\documentclass[preprint,aip,apl,amsmath,amssymb,superscriptaddress,12pt]{revtex4}

\usepackage{graphicx}
\usepackage{dcolumn}
\usepackage{bm,color}
\bibstyle{apsrev}
\usepackage{doublespace}

\begin{document}


\title{Atomic-scale visualization of initial growth of homoepitaxial
   SrTiO$_{\mathbf{3}}$ thin film on an atomically ordered substrate
}

\author{Ryota Shimizu}
\affiliation
{WPI-Advanced Institute for Materials Research, Tohoku University,
Sendai 980-8577, Japan}
\affiliation
{Department of Chemistry, The University of Tokyo,Tokyo 113-0033, Japan}
\author{Katsuya Iwaya}
\affiliation
{WPI-Advanced Institute for Materials Research, Tohoku University,
Sendai 980-8577, Japan}
\author{Takeo Ohsawa}
\affiliation{WPI-Advanced Institute for Materials Research, Tohoku University,
Sendai 980-8577, Japan}
\author{Susumu Shiraki}
\affiliation{WPI-Advanced Institute for Materials Research, Tohoku University,
Sendai 980-8577, Japan}
\author{Tetsuya Hasegawa}
\affiliation{Department of Chemistry, The University of Tokyo,Tokyo 113-0033, Japan}
\author{Tomihiro Hashizume}
\affiliation{WPI-Advanced Institute for Materials Research, Tohoku University,
Sendai 980-8577, Japan}
\affiliation{Central Research Laboratory, Hitachi, Ltd., Saitama 350-0395, Japan}
\affiliation{Department of Physics, Tokyo Institute of Technology,
Tokyo 152-8551, Japan}
\author{Taro Hitosugi}
\email{hitosugi@wpi-aimr.tohoku.ac.jp}
\affiliation{WPI-Advanced Institute for Materials Research, Tohoku University,
Sendai 980-8577, Japan}
\affiliation{PRESTO, Japan Science and Technology Agency, Kawaguchi,
Saitama 332-0012, Japan}


\begin{abstract}
  The initial homoepitaxial growth of SrTiO$_3$ on a ($\sqrt{13}\times\sqrt{13}$)-$R33.7^{\circ}$ SrTiO$_3$(001) substrate surface, which can be prepared under oxide growth conditions, is atomically resolved by scanning tunneling microscopy. The identical ($\sqrt{13}\times\sqrt{13}$) atomic structure is clearly visualized on the deposited SrTiO$_3$ film surface as well as on the substrate. This result indicates the transfer of the topmost Ti-rich ($\sqrt{13}\times\sqrt{13}$) structure to the film surface and atomic-scale coherent epitaxy at the film/substrate interface. Such atomically ordered SrTiO$_3$ substrates can be applied to the fabrication of atom-by-atom controlled oxide epitaxial films and heterostructures.
\end{abstract}
\maketitle

\section{Introduction}
Transition metal oxides (TMOs) have attracted significant interest due to various fascinating phenomena such as high-$T_{\mathrm{c}}$ superconductivity and colossal magnetoresistance, which originate from the interplay between spin, charge, orbital, and lattice degrees of freedom.\cite{RMP_70_1039,Science_288_462} Recently, the interfaces of TMO heterostructures have exhibited unique physical properties that are different from their constituent materials.\cite{Nature_427_423} To further explore multifunctional properties that utilize the electronic degrees of freedom, it is crucial to grow oxide thin films in an atom-by-atom manner by monitoring the initial stage of epitaxial behavior at the atomic level.\cite{JAmCeramSoc_91_2429}

The initial growth of TMO thin films has been typically characterized using diffraction methods, such as \textit{in situ} reflection high-energy electron diffraction (RHEED) measurements. Scanning tunneling microscopy (STM) provides information regarding real-space surface structures with high spatial resolution, as demonstrated in cuprate,\cite{JAP_75_5227,Science_267_71} manganite,\cite{PRL_95_237210,PRL_102_066104} and titanate\cite{APEX_3_075701,JAP_108_073710} thin films.
\textcolor{black}{Focusing on the initial stage of one- or two-unit-cell growth, the latest report showed selective island nucleation of TMOs by controlling surface reconstructions.\cite{APL_98_161908} However, atomic-scale understandings of initial thin film growth, involving the formation of the surface and interface, have been elusive,}
due to poor preparation processes of the substrate surfaces under actual oxide growth conditions.

In this article, we demonstrate atomic-scale visualization of the initial homoepitaxial SrTiO$_3$ growth process using STM. An annealing method was first developed to prepare a ($\sqrt{13}\times\sqrt{13}$)-$R$33.7$^{\circ}$ (hereafter described as ($\sqrt{13}\times\sqrt{13}$) for brevity) reconstructed SrTiO$_3$(001) surface under typical conditions for oxide thin film growth. A few monolayer (ML) SrTiO$_3$ thin films were subsequently deposited on the ($\sqrt{13}\times\sqrt{13}$) reconstructed surface.
\textcolor{black}{The topmost Ti-rich ($\sqrt{13}\times\sqrt{13}$) structures, consisting of an additional Ti-rich layer,\cite{PhysRevLett_106_176102}}
are clearly identified on the first- and second-layer surfaces of the film, which indicates atomic-scale coherent epitaxy between the thin film and substrate. Our study shifts the oxide thin film growth and nanostructure research, from unit-cell level to atomic level.

\section{Results and Discussion}
Figure 1(a) shows a widely accepted annealing procedure used to obtain wide terraces with equidistant steps of unit-cell height (step/terrace structure).\cite{Science_266_1540} The BHF-etched substrate with step/terrace structure was annealed under an oxygen partial pressure ($P_{\mathrm{O}_2}$) of $1\times10^{-6}$ Torr at 1000$^{\circ}$C for 30 min. A streaky RHEED pattern was confirmed and a clear step/terrace structure was observed. However, on a smaller scale, no ordered structure was found in the STM images [Fig. 1(c)].\cite{JAP_108_073710}

The annealing sequence was optimized to improve the atomic surface structure [Fig. 1(d)]. The samples were degassed at 500$^{\circ}$C for 2 h at $P_{\mathrm{O}_2}=1\times10^{-6}$ Torr, and then annealed at 850$^{\circ}$C for 40 min under $P_{\mathrm{O}_2}=1\times10^{-5}$ Torr to obtain large ($\sqrt{13}\times\sqrt{13}$) domains. During this annealing, the samples were heated to 1000$^{\circ}$C for 3 min to straighten the step structures. 
\textcolor{black}{The key of this annealing process is to suppress the oxygen deficiencies in bulk crystal. We found that the ($\sqrt{13}\times\sqrt{13}$) reconstruction did not appear when the substrate has many oxygen deficiencies,\cite{InPreparation} and thus we degassed the sample at as low temperature as possible, heated it at high temperature in a short time, and provided sufficient oxygen partial pressures.}
After annealing, a very sharp RHEED pattern was observed that corresponded to the ($\sqrt{13}\times\sqrt{13}$) reconstruction [Fig. 1(e)].\cite{PhysicaC_229_1} 
\textcolor{black}{This surface was previously reported using RHEED,\cite{PhysicaC_229_1} and STM,\cite{APL_96_231901} and more recently, an theoretical atomic arrangement of the surface structure was proposed using density functional theory calculations combined with transmission electron microscopy techniques.\cite{PhysRevLett_106_176102}}
In this model, the surface structure is composed of an additional Ti-rich ($\sqrt{13}\times\sqrt{13}$) layer formed on a bulk-like termination of a TiO$_2$ plane.\cite{PhysRevLett_106_176102} It should be stressed here that STM images also showed the
\textcolor{black}{single-phase}
($\sqrt{13}\times\sqrt{13}$) periodicity together with a clear step/terrace structure preferable for thin film deposition [Fig. 1(f)].
\textcolor{black}{Note that the bright spots observed at the $c$($\sqrt{13}\times\sqrt{13}$) sites are associated with oxygen vacancies,\cite{InPreparation} for example, reduced Ti states or OH impurities similarly reported on a rutile TiO$_2$(110) surface,\cite{SurfSci_598_226} because the density of bright spots is dependent on the oxygen partial pressure. Further studies on the bright-point defects are needed.}

To date, a large number of SrTiO$_3$ surface structures have been reported,\cite{SurfSci_505_1,SurfSci_542_177,SurfSciRep_62_431} whereas little attention has been paid to the stability of those surface structures under the film growth conditions, in particular, in an oxygen partial pressure at high temperatures. Considering our results and the previous report on the preparation of ($\sqrt{13}\times\sqrt{13}$) reconstructed surface in an external oxygen-flowing tube furnace,\cite{PhysicaC_229_1,PhysRevLett_106_176102} we anticipate this reconstructed surface to be stable in a wide range of $P_{\mathrm{O}_2}$, from UHV to 1 atm. Further, the surface was found to be robust at various temperatures, from room temperature to 850$^{\circ}$C, because the RHEED patterns were unchanged during cooling down to room temperature at a rate of 3-5$^{\circ}$C/s, representing no phase transition of the surface structure. In terms of the understanding of the surface stability as well as the atomic arrangement of the structure, the ($\sqrt{13}\times\sqrt{13}$) reconstructed surface is an appropriate substrate to fabricate oxide thin films and heterointerfaces.

In order to fabricate higher-grade heterostructures, it is also particularly important to grow well-defined epitaxial films on atomically ordered substrates. We next optimized the laser fluence for SrTiO$_3$ homoepitaxial growth on the ($\sqrt{13}\times\sqrt{13}$) SrTiO$_3$ substrates. Figure 2(a) shows out-of-plane X-ray diffraction (XRD) results acquired around (004) peak of the 100-nm thick SrTiO$_3$ films. Except for the film deposited at 0.42 J/cm$^2$, the XRD peaks of (004) appear at lower 2$\theta$ values than the substrate peak 2$\theta=104.153^{\circ}$, indicating lattice expansion along the $c$ axis, which can be interpreted as the deposition of nonstoichimetric films.\cite{APL_87_241919} Using this optimal value, 0.42 J/cm$^2$, a clear RHEED intensity oscillation was observed in a layer-by-layer manner and persisted until stopping deposition (Fig. 2(b)). Therefore, this deposition condition assures the well-controlled epitaxial growth of nearly stoichiometic SrTiO$_3$.

The application of the atomically defined substrate to the conventional oxide epitaxial growth allows for atomic-scale visualization of the initial growth process near the interface between the SrTiO$_3$ layer and the substrate. 
\textcolor{black}{In the pulsed laser deposition (PLD) process, an ablated target material is vaporized to form a plume, where each particle has a kinetic energy of around several tens of electronvolts, and then reaches a substrate. The atoms adsorbed on the SrTiO$_3$ surface diffuse and finally a nucleation of islands occurs.\cite{PLDBook,PLDBook2,ProgSurfSci_76_163}}
Some of possible growth processes for SrTiO$_3$ deposition on the well-defined ($\sqrt{13}\times\sqrt{13}$) surface are illustrated in Fig. 3. For example, 
\textcolor{black}{(a) the excess Ti in the ($\sqrt{13}\times\sqrt{13}$) layer is embedded at the interface. For (b), as a result of the diffusion of the excess Ti to the substrate and the film, an interdiffusion region\cite{Book_SurfSci} is formed near the interface, and a different atomic arrangement appears on the film surface. For (c), the ($\sqrt{13}\times\sqrt{13}$) structure is transferred to the film surface, which is analogous to the homoepitaxial growth of GaAs.\cite{GaAs_review}}


To determine which growth process in Fig. 3 occurs, we show, in Fig. 4, STM images of 0.3 ML and 1.6 ML SrTiO$_3$ films grown on a ($\sqrt{13}\times\sqrt{13}$) reconstructed surface. The exact thickness was evaluated from island-coverage in wide-scale STM images (Fig. 4(a),(b) inset). A ($\sqrt{13}\times\sqrt{13}$)-based mesh structure was clearly evident, not only on the substrate, but also on the first layer of SrTiO$_3$ islands. Growth of a 1.6 ML SrTiO$_3$ film on a ($\sqrt{13}\times\sqrt{13}$) reconstructed substrate surface was also observed and the ($\sqrt{13}\times\sqrt{13}$) structure was identified on the second layer of SrTiO$_3$, as indicated in Fig. 4(b). These results support the third mechanism given in Fig. 3, in which the atomic structure of the substrate surface is transferred to the film surface, and therefore coherent epitaxy of SrTiO$_3$ can be realized at the interface between the thin film and substrate.

\textcolor{black}{The atomically-ordered ($\sqrt{13}\times\sqrt{13}$) substrate is applicable to further atomic-scale interfacial studies. For example, at the LaAlO$_3$/SrTiO$_3$ interface, where the physical properties depend on the oxygen partial pressures during growth, it is of great importance to understand the initial growth behavior under several growth conditions. Due to the robustness in a wide range of oxygen partial pressures as mentioned above, the ($\sqrt{13}\times\sqrt{13}$) substrate is a good candidate as the starting point to investigate fascinating properties emerging at various TMO heterointerfaces. In addition, the SrTiO$_3$(001) reconstructed surface has been used as a template to control novel nanostructures of metals and molecules.\cite{PhysRevLett_94_046103,ChemCommun_2941_2007,Nanotechnology_18_075301} Two attractive features of the ($\sqrt{13}\times\sqrt{13}$) surface (two-dimensional square lattice and large lattice parameter of around 1.4 nm) have potential to assemble large molecules such as phthalocyanine in an epitaxial manner.}

\section{Conclusions}
In summary, the initial growth process of homoepitaxial SrTiO$_3$ thin film was investigated at the atomic level using STM. An annealing process was first developed to obtain a ($\sqrt{13}\times\sqrt{13}$)-reconstructed SrTiO$_3$(001) substrate that is stable under typical conditions for oxide thin film deposition. After the homoepitaxial growth of SrTiO$_3$ on the ($\sqrt{13}\times\sqrt{13}$) reconstructed surface, ($\sqrt{13}\times\sqrt{13}$) structures were observed, even on the first and second layer of the SrTiO$_3$ thin films, which demonstrates the atomic-scale coherent epitaxy between thin film and substrate. This atomically controlled substrate surface has the potential to replace the conventional "step-and-terrace surface" for the fabrication of higher quality thin films and heterostructures. The application of this surface to heteroepitaxy would open up atomic-scale investigations of oxide thin film growth and studies of local physical properties of multifunctional oxide materials.

\section{Methods}
Niobium-doped (0.1 atom \%) SrTiO$_3$(001) single crystals (Shinkosha Corp.) were used as substrates to ensure conductivity for low-temperature scanning tunneling microscopy (STM) measurements. The ($\sqrt{13}\times\sqrt{13}$)-$R33.7^{\circ}$ reconstructed surfaces with step and terrace structures were prepared by buffered hydrofluoric acid (BHF) etching and subsequent annealing. Non-doped SrTiO$_3$ thin films were homoepitaxially grown on the ($\sqrt{13}\times\sqrt{13}$)-$R33.7^{\circ}$ substrate surfaces using pulsed laser deposition (PLD). The films were grown under an oxygen partial pressure ($P_{\mathrm{O}_2}$) of $1\times10^{-6}$ Torr at 700$^{\circ}$C by direct current resistive heating through the samples. A KrF excimer laser (wavelength of 248 nm) with a repetition rate of 2 Hz was employed, and the laser fluence at the target surface was 0.42 J/cm$^2$. The film growth was monitored from the \textit{in situ} RHEED intensity oscillation in a layer-by-layer growth manner. After growth, the samples were cooled to room temperature at a rate of 3-5$^{\circ}$C/s under $P_{\mathrm{O_2}} = 1\times10^{-6}$ Torr followed by immediate transfer to the STM without exposure of the sample surface to air.\textcolor{black}{\cite{RevSciInstrum_inpress}}
All STM measurements were conducted at 78 K.

\section{Acknowledgments}
This work was supported by the World Premier International Research Center Initiative (WPI Initiative), the Global COE Program for Chemistry Innovation through Cooperation of Science and Engineering, a Grant-in-Aid for Scientific Research on Priority Areas "Nano Materials Science for Atomic-scale Modification 474" from the Ministry of Education, Culture, Sports, Science and Technology (MEXT), Japan, and the A3 Foresight Program of the Japanese Society for the Promotion of Science (JSPS). 


\newpage
\noindent
{\bf{\large Figure captions}} \vspace{3mm} \\
Figure 1: \\
(a) Schematic diagram of the typical annealing process to obtain the step/terrace surface. (b) RHEED pattern and (c) STM image of the SrTiO$_3$(001) substrate after the typical annealing process shown in (a) (20$\times$20 nm$^2$, sample-bias-voltage $V_{\mathrm{s}} = +1.5$ V, and set-point tunneling current $I_{\mathrm{t}} = 30$ pA). The inset shows a wide-view STM image (400$\times$400 nm$^2$). (d) Schematic diagram of the optimized annealing process for the ($\sqrt{13}\times\sqrt{13}$) surface. (e) RHEED pattern, and (f) STM image of the SrTiO$_3$(001) substrate after the optimized annealing process shown in (d). The scan size and the scanning conditions are the same as those used for (c). Each RHEED pattern was obtained at room temperature along the [010] azimuth.
\\ \phantom{o} \vspace{-4mm}\\
Figure 2: \\
(a): Out-of-plane XRD results obtained around the (004) peak of the SrTiO$_3$ thin films deposited with various laser fluences. (b): RHEED intensity oscillation of the specular reflected beam monitored at the beginning and the end of the deposition. The clear intensity oscillation continued until stop deposition with the 254 unit-cell thickness($\approx$100 nm).
\\ \phantom{o} \vspace{-4mm} \\
Figure 3: \\
Schematic illustration of possible homoepitaxial growth mechanisms of SrTiO$_3$ on the ($\sqrt{13}\times\sqrt{13}$) substrate.
\textcolor{black}{The Ti-rich ($\sqrt{13}\times\sqrt{13}$) layer is (a) embedded at the interface and a new structure is formed on the thin film, (b) diffused to the substrate and the film, forming interdiffusion region and a different surface structure appears on the thin film surface, (c) transferred to the film surface.}
\\ \phantom{o} \vspace{-4mm}\\
Figure 4: \\
STM images of ultrathin (a) 0.3 ML, and (b) 1.6 ML SrTiO$_3$(001) films grown on ($\sqrt{13}\times\sqrt{13}$) substrate surfaces (20$\times$20 nm$^2$, $V_{\mathrm{s}} = +1.5$ V, $I_{\mathrm{t}} = 30$ pA). The ($\sqrt{13}\times\sqrt{13}$) unit cell is indicated by dotted lines. The insets show wide-scale STM images used to estimate the coverage of the thin films (100$\times$100 nm$^2$).
\newpage

\providecommand*\mcitethebibliography{\thebibliography}
\csname @ifundefined\endcsname{endmcitethebibliography}
  {\let\endmcitethebibliography\endthebibliography}{}

\newpage
\noindent
{\bf{\large Figures}} \vspace{3mm}
\begin{center}
\includegraphics[width=120mm,bb=0 0 300 340]{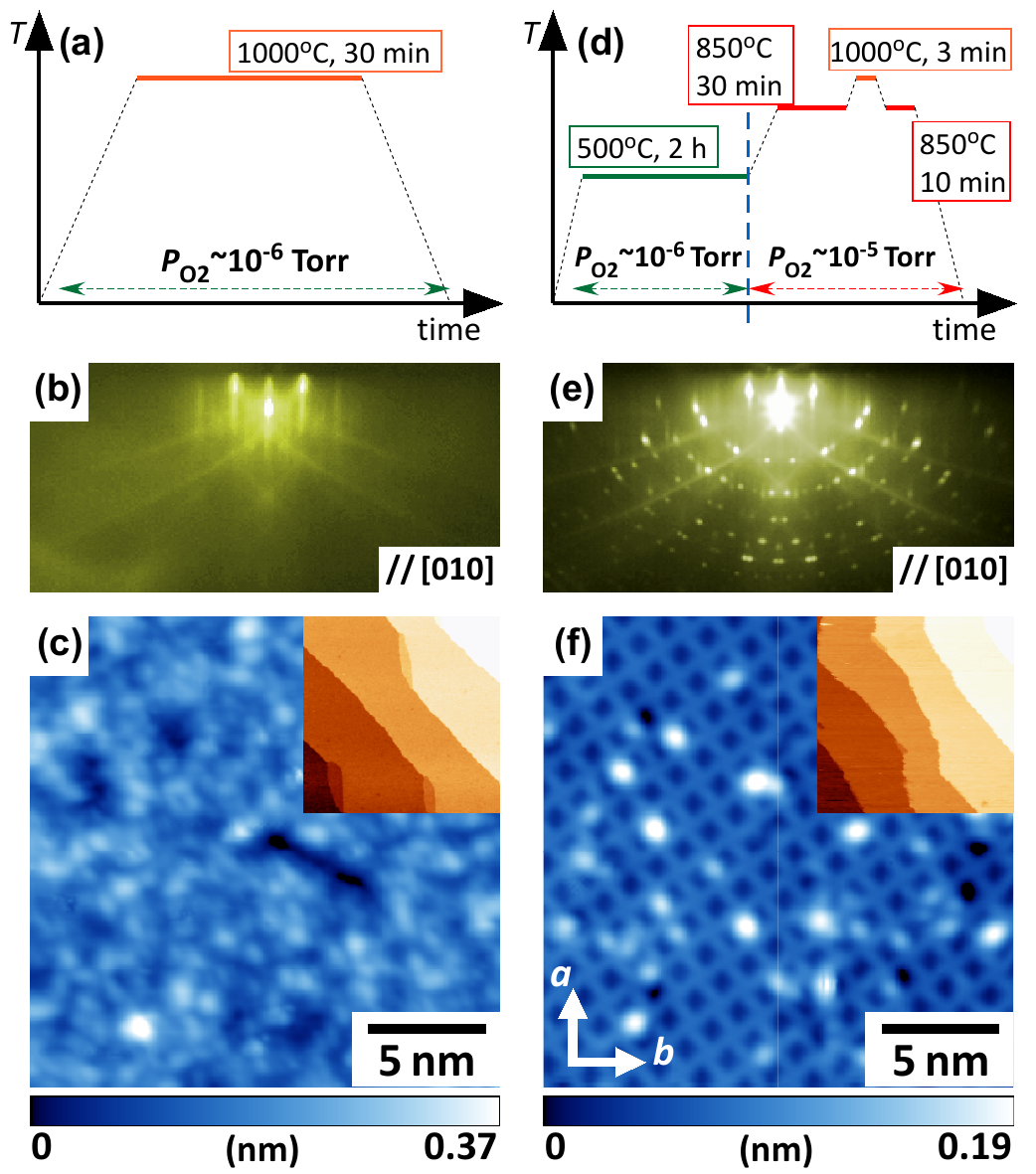} \vspace{5mm} \\
Figure 1 (a)-(f)
\end{center}
\newpage
\begin{center}
\includegraphics[width=150mm,bb=0 0 623 396]{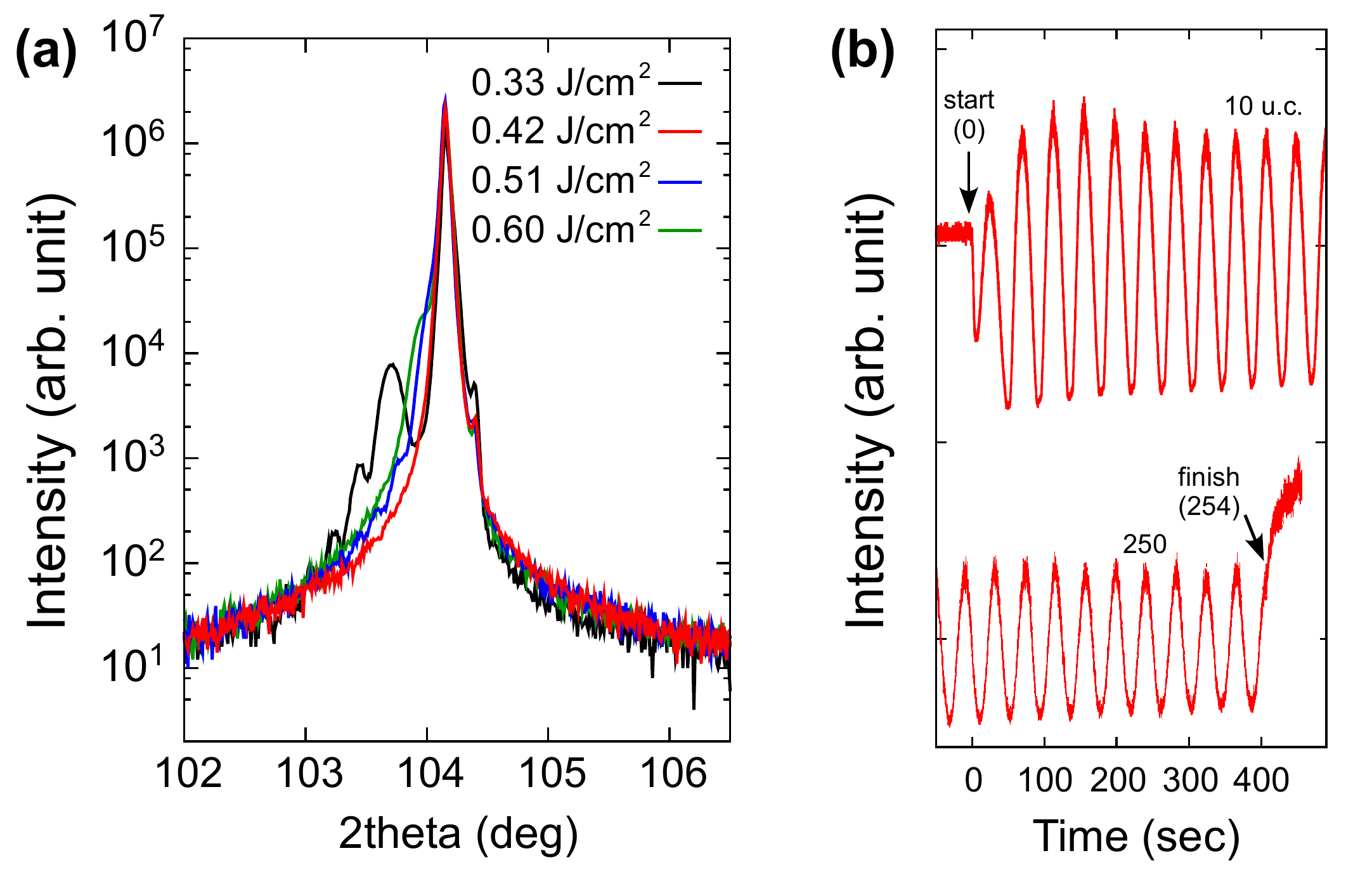} \vspace{5mm} \\
Figure 2 (a)-(b)
\end{center}
\newpage
\begin{center}
\includegraphics[width=150mm,bb=0 0 425 255]{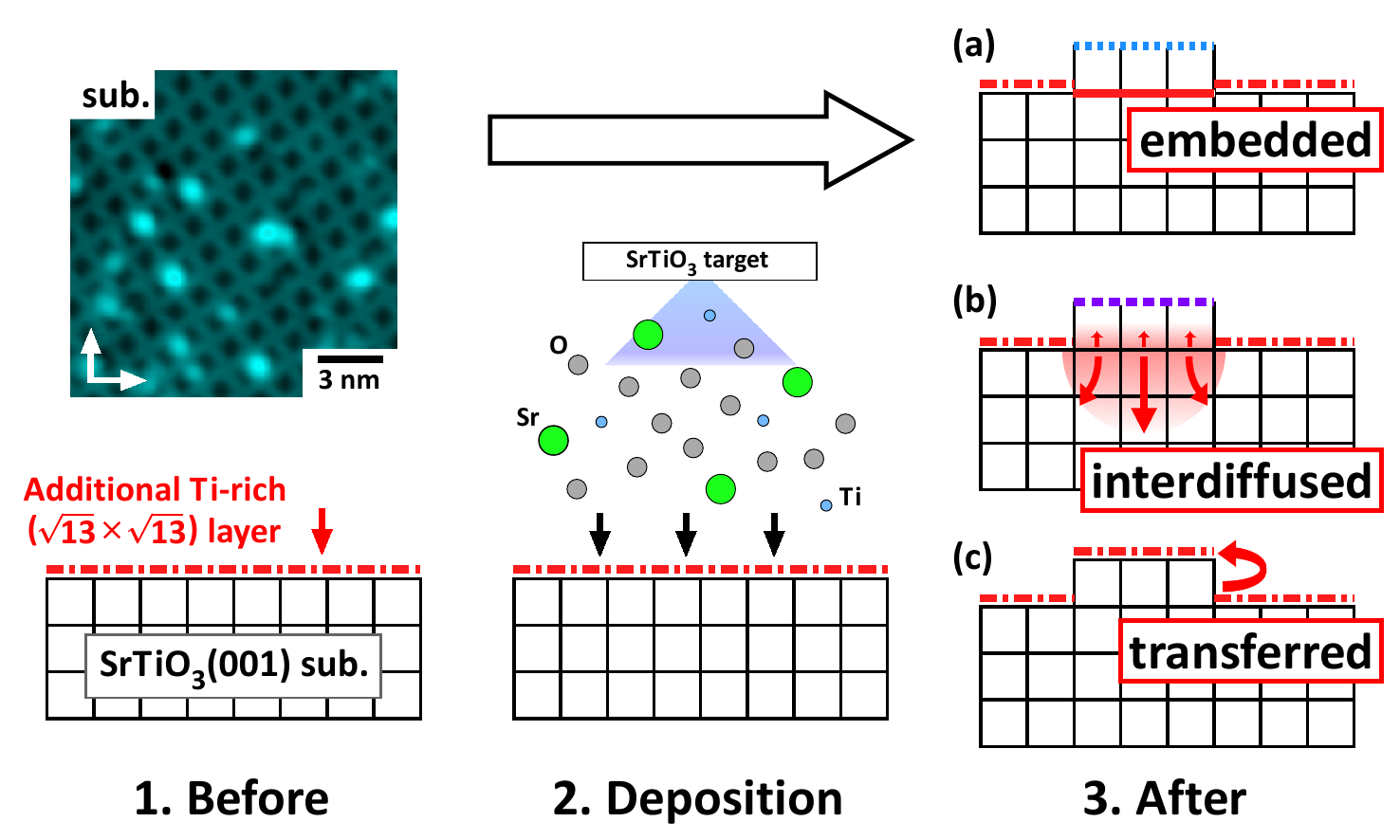} \vspace{5mm} \\
Figure 3 (a)-(c)
\end{center}
\newpage
\begin{center}
\includegraphics[width=150mm,bb=0 0 481 265]{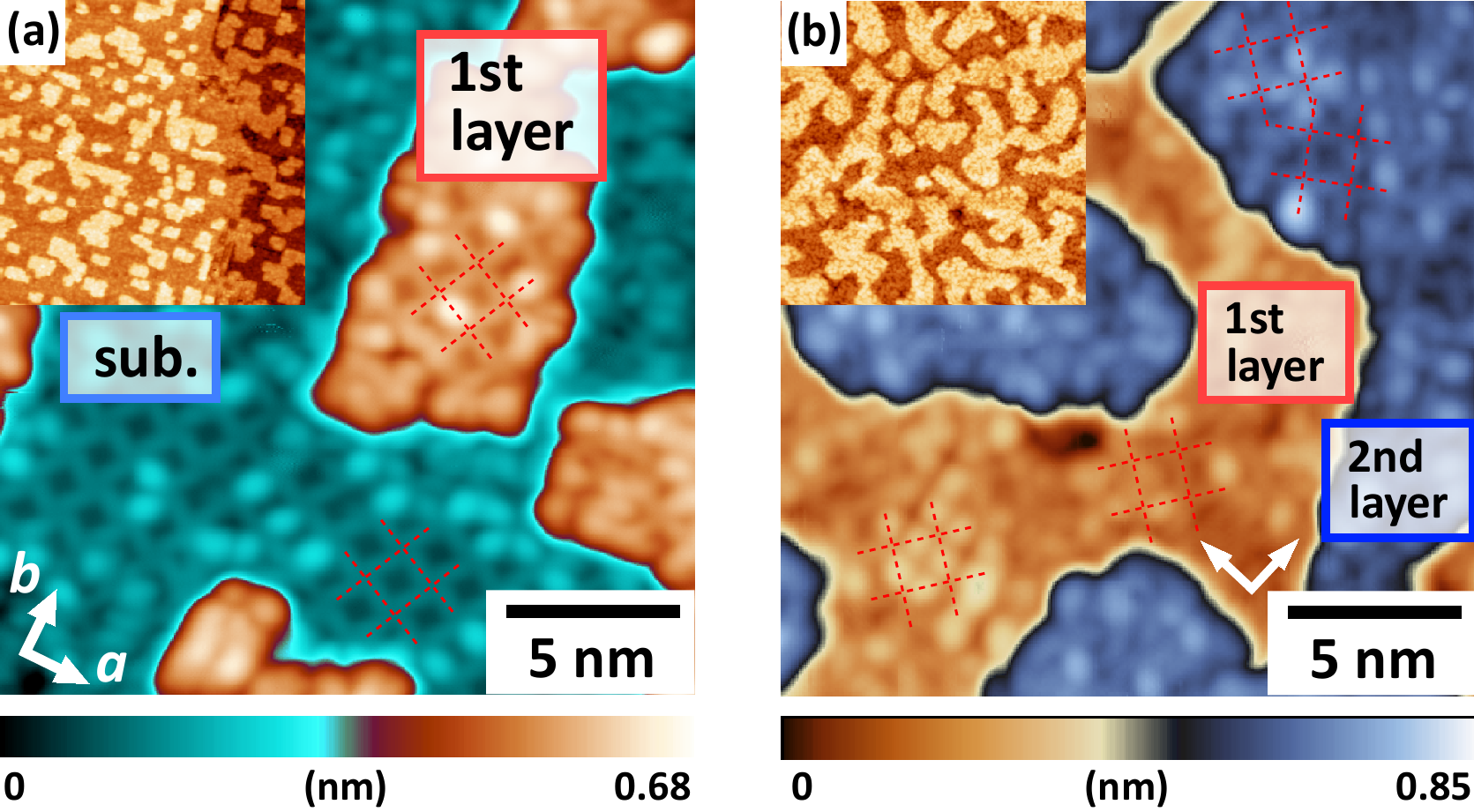} \vspace{5mm} \\
Figure 4 (a)-(b)
\end{center}
\end{document}